\documentclass{ws-ijmpa}

\begin{document}

\markboth{Zhi-zhong Xing, He Zhang and Shun Zhou} {Generalized
Friedberg-Lee model and CP violation from $\mu-\tau$ symmetry
breaking}

%
\catchline{}{}{}{}{}
%

\title{Generalized Friedberg-Lee model for neutrino masses and leptonic
CP violation from $\mu$-$\tau$ symmetry breaking}

\author{Zhi-zhong Xing, He Zhang and Shun Zhou}

\address{Institute of High Energy Physics, Chinese Academy of Sciences,
Beijing 100049, China}

\maketitle

\begin{history}
\received{\today}
\end{history}

\begin{abstract}
Assuming the Majorana nature of massive neutrinos, we generalize
the Friedberg-Lee neutrino mass model to include CP violation in
the neutrino mass matrix $M^{}_\nu$. The most general case with
all the free parameters of $M^{}_\nu$ being complex is discussed.
We show that a favorable neutrino mixing pattern (with
$\theta^{}_{12} \approx 35.3^\circ$, $\theta^{}_{23} = 45^\circ$,
$\theta^{}_{13} \neq 0^\circ$ and $\delta = 90^\circ$) can
naturally be derived from $M^{}_\nu$, if it has an approximate or
softly-broken $\mu$-$\tau$ symmetry. We also point out a different
way to obtain the nearly tri-bimaximal neutrino mixing pattern
with $\delta = 0^\circ$ and non-vanishing Majorana phases.

\end{abstract}
\ccode{PACS numbers: 14.60.Lm; 14.60.Pq; 95.85.Ry}

\vspace{0.2cm}

Recently, a novel neutrino mass model has been proposed by Friedberg
and Lee (FL).\cite{Lee} The neutrino mass operator in the FL model
is simply given by
\begin{eqnarray}
{\cal L}^{}_{\nu- \rm mass} & = & a \left (\overline{\nu}^{}_\tau -
\overline{\nu}^{}_\mu \right ) \left (\nu^{}_\tau - \nu^{}_\mu
\right ) + b \left (\overline{\nu}^{}_\mu - \overline{\nu}^{}_e
\right ) \left (\nu^{}_\mu - \nu^{}_e \right ) + c \left
(\overline{\nu}^{}_e - \overline{\nu}^{}_\tau \right ) \left
(\nu^{}_e - \nu^{}_\tau \right )
\nonumber \\
& & + m^{}_0 \left (\overline{\nu}^{}_e \nu^{}_e +
\overline{\nu}^{}_\mu \nu^{}_\mu + \overline{\nu}^{}_\tau
\nu^{}_\tau \right ) \; ,
\end{eqnarray}
where the parameters $a$, $b$, $c$ and $m^{}_0$ are all assumed to
be {\it real}, and the charged-lepton mass matrix is taken to be
diagonal. A salient feature of ${\cal L}^{}_{\nu- \rm mass}$ is its
partial gauge-like symmetry; i.e., its $a$, $b$ and $c$ terms are
invariant under the transformation $\nu^{}_\alpha \rightarrow
\nu^{}_\alpha + z$ (for $\alpha = e, \mu, \tau$) with $z$ being a
space-time independent constant element of the Grassmann
algebra.\cite{Lee} From Eq. (1), one can directly write down the
neutrino mass matrix:
\begin{equation}
M^{}_\nu \; = \; m^{}_0 \left ( \begin{matrix}1 & 0 & 0 \cr 0 & 1 &
0 \cr 0 & 0 & 1 \cr\end{matrix} \right ) ~ + ~ \left (
\begin{matrix} b+c & -b & -c \cr -b & ~ a+b ~ & -a \cr -c & -a & c+a
\cr\end{matrix} \right ) \; .
\end{equation}
Two interesting features can be inferred from the diagonalization of
$M^{}_\nu$. First, the neutrino mass matrix takes a magic
form,\cite{Magic} in which the sums of rows and columns are all
equal to $m^{}_0$. The unitary matrix used to diagonalize $M^{}_\nu$
must have one eigenvector with three equal components $1/\sqrt{3}~$.
Second, when $b=c$ holds, it is very easy to check that the neutrino
mass operator ${\cal L}^{}_{\nu- \rm mass}$ has the exact
$\mu$-$\tau$ symmetry (i.e., ${\cal L}^{}_{\nu- \rm mass}$ is
invariant under the exchange of $\mu$ and $\tau$
indices).\cite{Symmetry} In addition, one may consider to remove one
degree of freedom from ${\cal L}^{}_{\nu- \rm mass}$ or $M^{}_\nu$
(for instance, by setting $c =0$).\cite{Lee}

To include CP or T violation into the FL model, one may insert the
phase factors $e^{\pm i \eta}$ into Eq. (1) by replacing the term $c
\left (\overline{\nu}^{}_e - \overline{\nu}^{}_\tau \right ) \left
(\nu^{}_e - \nu^{}_\tau \right )$ with the term $c \left (e^{-i\eta}
\overline{\nu}^{}_e - \overline{\nu}^{}_\tau \right ) \left
(e^{+i\eta} \nu^{}_e - \nu^{}_\tau \right )$.\cite{Lee} The
resultant neutrino mass matrix is no longer symmetric, hence it
describes Dirac neutrinos instead of Majorana neutrinos. However, in
most of the realistic models, the Majorana nature is preferable to
the Dirac nature of neutrinos. Hence, in this work, we aim to
generalize the FL model to include CP and T violation for massive
Majorana neutrinos.

Let us start from the generic analysis with all the parameters of
$M^{}_\nu$ in Eq. (2) being complex. For Majorana neutrinos,
$M^{}_\nu$ is symmetric and can be diagonalized by the
transformation $V^\dagger M^{}_\nu V^* = {\rm Diag} \{ m^{}_1,
m^{}_2, m^{}_3 \}$, in which $m^{}_i$ (for $i=1,2,3$) stand for
the neutrino masses. After a straightforward calculation, the
neutrino mixing matrix $V$ turns out to be
\begin{equation}
V = \left ( \begin{matrix} \displaystyle \frac{2}{\sqrt{6}}
\cos\frac{\theta}{2} & \displaystyle \frac{1}{\sqrt{3}} &
\displaystyle \frac{2}{\sqrt{6}} \sin\frac{\theta}{2} e^{-i\delta}
\cr\cr \displaystyle -\frac{1}{\sqrt{6}} \cos\frac{\theta}{2} -
\frac{1}{\sqrt{2}} \sin\frac{\theta}{2} e^{i\delta} &
\displaystyle ~~ \frac{1}{\sqrt{3}} ~~ & \displaystyle
\frac{1}{\sqrt{2}} \cos\frac{\theta}{2} - \frac{1}{\sqrt{6}}
\sin\frac{\theta}{2} e^{-i\delta} \cr\cr \displaystyle
-\frac{1}{\sqrt{6}} \cos\frac{\theta}{2} + \frac{1}{\sqrt{2}}
\sin\frac{\theta}{2} e^{i\delta} & \displaystyle
\frac{1}{\sqrt{3}} & \displaystyle -\frac{1}{\sqrt{2}}
\cos\frac{\theta}{2} - \frac{1}{\sqrt{6}} \sin\frac{\theta}{2}
e^{-i\delta} \cr\end{matrix}  \right ) \left (
\begin{matrix} e^{i\rho} & 0 & 0 \cr\cr 0 & e^{i\sigma} & 0 \cr\cr 0
& 0 & 1 \cr\end{matrix} \right ) ,
\end{equation}
where the explicit expressions of $\theta$ and $\delta$ are
\begin{eqnarray}
\tan \theta  =  \frac{\sqrt{X^2 + Y^2}}{Z} \ , \ \ \ \ \ \tan\delta
=  \frac{X}{Y} \; ,
\end{eqnarray}
and the definitions of $X$, $Y$ and $Z$ can be found in Ref.
\refcite{FL}. Furthermore, three mass eigenvalues of $M^{}_\nu$
and two Majorana phases of $V$ are found to be
\begin{eqnarray}
m^{}_1 & = & \left | T^{}_1 \cos^2\frac{\theta}{2} - T_2\sin\theta
e^{-i\delta} + T^{}_3\sin^2\frac{\theta}{2} e^{-i 2\delta} \right |
\; , \ \ \ \ \ m^{}_2 = \left |m^{}_0 \right | \; ,
\nonumber \\
m^{}_3 & = & \left | T^{}_3 \cos^2\frac{\theta}{2} +
T^{}_2\sin\theta e^{+i\delta} + T^{}_1\sin^2\frac{\theta}{2} e^{+i
2\delta} \right | \; ;
\end{eqnarray}
and
\begin{eqnarray}
\rho & = & \frac{1}{2} \arg \left( \frac{\displaystyle T^{}_1
\cos^2\frac{\theta}{2} - T^{}_2\sin\theta e^{-i\delta} +
T^{}_3\sin^2\frac{\theta}{2} e^{-i 2\delta}}{\displaystyle T^{}_3
\cos^2\frac{\theta}{2} + T^{}_2\sin\theta e^{+i\delta} +
T^{}_1\sin^2\frac{\theta}{2} e^{+i 2\delta}} \right) \; ,
\nonumber \\
\sigma & = & \frac{1}{2}\arg \left( \frac{m^{}_0}{\displaystyle
T^{}_3 \cos^2\frac{\theta}{2} + T^{}_2\sin\theta e^{+i\delta} +
T^{}_1\sin^2\frac{\theta}{2} e^{+i 2\delta}} \right) \; .
\end{eqnarray}
Again, the expressions of $T^{}_i$ have also been listed in Ref.
\refcite{FL}.

We proceed to consider two special but interesting scenarios of the
generalized FL model and explore their respective consequences on
three neutrino mixing angles and three CP-violating phases.

Scenario (A): $a$ and $m^{}_0$ are real, and $b = c^*$ are
complex. Note that the $\mu$-$\tau$ symmetry of $M^{}_\nu$ is
softly broken in this case, because $|b|=|c|$ holds. By using the
generic results given in Eqs. (3)-(6), one can easily arrive at
$\tan\theta ={\sqrt{3} ~ {\rm Im} \left (b \right )} /\left [
m^{}_0 + a + 2 {\rm Re} \left (b \right ) \right ]$,
$\delta=90^\circ$,
\begin{eqnarray}
m^{}_1 & = &  \sqrt{\left [m^{}_0 + a + 2{\rm Re} \left (b \right )
\right ]^2 + 3 \left [ {\rm Im} \left (b \right ) \right ]^2} ~ - a
+ {\rm Re} \left (b \right ) \; , \ \ \ \ \ m^{}_2  =  m^{}_0 \; ,
\nonumber \\
m^{}_3 & = & \sqrt{\left [m^{}_0 + a + 2{\rm Re} \left (b \right )
\right ]^2 + 3 \left [ {\rm Im} \left (b \right ) \right ]^2} ~ +
a - {\rm Re} \left (b \right ) \; ,
\end{eqnarray}
together with $\rho=\sigma=0$. Comparing our results with the
well-known standard parametrization,\cite{SP} we immediately obtain
$\sin\theta^{}_{12} = 1/\left({\sqrt{2 + \cos\theta}}\right)$, $
\sin\theta^{}_{23}  = 1/{\sqrt{2}} ~$, and
 $\sin\theta^{}_{13}  =  2/\left[{\sqrt{6}}
 \sin({\theta}/{2})\right]$. The leptonic Jarlskog parameter $\cal J$, which
is a rephasing-invariant measure of CP violation in neutrino
oscillations,\cite{J} reads ${\cal J} = \sin\theta /(6\sqrt{3})$. If
$\theta =0^\circ$ holds, the tri-bimaximal neutrino mixing pattern
(with $\tan\theta^{}_{12} = 1/\sqrt{2}$ or $\theta^{}_{12} \approx
35.3^\circ$, $\theta^{}_{23} = 45^\circ$ and $\theta^{}_{13} =
0^\circ$)\cite{TB} will be reproduced. One can see that the soft
breaking of $\mu$-$\tau$ symmetry leads to both $\theta^{}_{13} \neq
0^\circ$ and ${\cal J} \neq 0$, but it does not affect the favorable
result $\theta^{}_{23} = 45^\circ$ given by the tri-bimaximal mixing
pattern. On the other hand, $\sin\theta^{}_{12} \approx 1/\sqrt{3}$
is an excellent approximation, since $\theta$ must be small to
maintain the smallness of $\theta^{}_{13}$. In view of
$\theta^{}_{13} < 10^\circ$,\cite{Vissani} we obtain $\theta
\lesssim 24.6^\circ$ and ${\cal J} \lesssim 0.04$. It is possible to
measure ${\cal J} \sim {\cal O}(10^{-2})$ in the future
long-baseline neutrino oscillation experiments. The neutrino masses
in scenario (A) rely on four real model parameters $m^{}_0$, $a$,
${\rm Re}(b)$ and ${\rm Im}(b)$. Thus it is easy to fit the neutrino
mass-squared differences $\Delta m^2_{21} = (7.2\ldots8.9)\times
10^{-5} ~ {\rm eV^2}$ and $\Delta m^2_{32} = \pm
(2.1\ldots3.1)\times 10^{-3} ~ {\rm eV^2}$.\cite{Vissani} Such a fit
should not involve any fine-tuning, because (a) the number of free
parameters is larger than the number of constraint conditions and
(b) three neutrino masses have very weak correlation with three
mixing angles. A detailed numerical analysis can be found in Ref.
\refcite{FL}, and a remarkable feature is that only the normal
neutrino mass hierarchy ($m^{}_1 < m^{}_2 < m^{}_3$) is allowed in
this scenario.

Scenario (B): $a$, $b$ and $c$ are all real, but $m^{}_0$ is
complex. By using Eqs. (3)-(6), we obtain $ \tan\theta = \sqrt{3}
\left (b - c \right )/{\left (b + c -2a \right ) }$, and
\begin{eqnarray}
m^{}_1 & = & \sqrt{\left [m^{}_{-} + {\rm Re} \left (m^{}_0 \right
) \right ]^2 + \left [ {\rm Im} \left ( m^{}_0 \right ) \right
]^2} \;\; , \ \ \ \ \
m^{}_2 =  \left |m^{}_0 \right | \; , \nonumber \\
m^{}_3 & = & \sqrt{\left [m^{}_{+} + {\rm Re} \left (m^{}_0 \right
) \right ]^2 + \left [ {\rm Im} \left ( m^{}_0 \right ) \right
]^2} \;\; ,
\end{eqnarray}
where $m^{}_{\pm} \; = \; \left ( a + b + c \right ) \pm \sqrt{a^2
+ b^2 + c^2 - ab - ac - bc} ~$. Two Majorana phases $\rho$ and
$\sigma$ are given by
\begin{eqnarray}
\tan 2 \rho & = & \frac{\left ( m^{}_+ - m^{}_- \right ) {\rm Im}
\left ( m^{}_0 \right )}{m^2_0 + m^{}_+ m^{}_- + \left ( m^{}_+ +
m^{}_- \right ) {\rm Re} \left ( m^{}_0 \right )} \; ,
\nonumber \\
\tan 2 \sigma & = & \frac{m^{}_+ {\rm Im} \left ( m^{}_0 \right
)}{m^2_0 + m^{}_+ {\rm Re} \left ( m^{}_0 \right )} \; .
\end{eqnarray}
Although $\rho$ and $\sigma$ have nothing to do with the behaviors
of neutrino oscillations, they may significantly affect the
neutrinoless double-beta decay.\cite{Xing02} Comparing our formulae
with the standard parametrization, we arrive at $\sin\theta^{}_{12}
= {1}/{\sqrt{2 + \cos\theta}}$, $\sin\theta^{}_{23}  =  {\sqrt{2 +
\cos\theta - \sqrt{3} \sin\theta}}/{\sqrt{2 \left (2 + \cos\theta
\right )}}$, $\sin\theta^{}_{13}  =  {2}/{\sqrt{6}}
\sin({\theta}/{2})$ together with $\delta = 0^\circ$ for the Dirac
phase of CP violation. The results for $\theta^{}_{12}$ and
$\theta^{}_{13}$ in this scenario are the same as those obtained in
scenario (A), but the Jarlskog parameter $\cal J$ is now vanishing.
Because of the $\mu$-$\tau$ symmetry breaking, $\theta^{}_{23}$ may
somehow deviate from the favorable value $\theta^{}_{23} =
45^\circ$. Given $\theta \lesssim 24.6^\circ$ corresponding to
$\theta^{}_{13} < 10^\circ$, $\theta^{}_{23}$ is allowed to vary in
the range $37.8^\circ \lesssim \theta^{}_{23} \leq 45^\circ$. The
neutrino masses depend on five real model parameters $a$, $b$, $c$,
${\rm Re}(m^{}_0)$ and ${\rm Im}(m^{}_0)$. Hence there is sufficient
freedom to fit two observed neutrino mass-squared differences
$\Delta m^2_{21}$ and $\Delta m^2_{32}$. Our careful numerical
analysis, which has been done in Ref.~\refcite{FL}, shows that both
normal ($m^{}_1 < m^{}_2 < m^{}_3$) and inverted ($m^{}_3 < m^{}_1 <
m^{}_2$) neutrino mass hierarchies are allowed, and the Majorana
phases ($\rho$, $\sigma$) are less restricted in scenario (B).

Although our discussions about the generalized FL model are
restricted to low-energy scales, it can certainly be extended to a
superhigh-energy scale (e.g., the GUT scale or the seesaw scale). In
this case, one should take into account the radiative corrections to
both neutrino masses and flavor mixing parameters when they run from
the high scale to the electroweak scale.\cite{Luo}

We conclude that the $\mu$-$\tau$ symmetry and its slight breaking
are useful and suggestive for model building. We expect that a
stringent test of the generalized FL model, in particular its two
simple and instructive scenarios, can be achieved in the near future
from the neutrino oscillation and neutrinoless double-beta decay
experiments.

\section*{Acknowledgments}
This work is supported in part by the National Natural Science
Foundation of China.




\begin{thebibliography}{99}

\bibitem{Lee} R. Friedberg and T. D. Lee, \emph{High Energy Phys. Nucl.
Phys.} {\bf 30}, 591 (2006).

\bibitem{Magic} C. S. Lam, \emph{Phys. Lett. B} {\bf 640}, 260 (2006).

\bibitem{Symmetry} See, e.g., T. Fukuyama and H. Nishiura, hep-ph/9702253;
R. N. Mohapatra and S. Nussinov, \emph{Phys. Rev. D} {\bf 60},
013002 (1999); Z. Z. Xing, \emph{Phys. Rev. D} {\bf 61}, 057301
(2000); \emph{Phys. Rev. D} {\bf 64}, 093013 (2001); \emph{Phys.
Rev. D} {\bf 74}, 013010 (2006).

\bibitem{FL} Z. Z. Xing, H. Zhang and S. Zhou, \emph{Phys. Lett. B} {\bf
641}, 189 (2006).

\bibitem{SP} Z. Z. Xing, \emph{Int. J. Mod. Phys. A} {\bf 19}, 1 (2004).

\bibitem{J} C. Jarlskog, \emph{Phys. Rev. Lett.} {\bf 55}, 1039 (1985);
D. D. Wu, \emph{Phys. Rev. D} {\bf 33}, 860 (1986).

\bibitem{TB} P. F. Harrison, D. H. Perkins and W. G. Scott, \emph{Phys.
Lett. B} {\bf 530}, 167 (2002); Z. Z. Xing, \emph{Phys. Lett. B}
{\bf 533}, 85 (2002).

\bibitem{Vissani} A. Strumia and F. Vissani, hep-ph/0606054.

\bibitem{Xing02} Z. Z. Xing, \emph{Phys. Rev. D} {\bf 64}, 093013 (2001);
\emph{Phys. Rev. D} {\bf 65}, 077302 (2002).

\bibitem{Luo} S. Luo, J. W. Mei and Z. Z. Xing, \emph{Phys. Rev. D} {\bf 72}, 053014
(2005); S. Luo and Z. Z. Xing, \emph{Phys. Lett. B} {\bf 632}, 341
(2006); \emph{Phys. Lett. B} {\bf 637}, 279 (2006); Z. Z. Xing and
H. Zhang, \emph{Commun. Theor. Phys.} {\bf 48}, 525 (2007).

\end{thebibliography}
\end{document}